\documentclass[fleqn,10pt]{wlscirep}
\usepackage[utf8]{inputenc}
\usepackage[T1]{fontenc}
\newcommand{\angstrom}{\mbox{\normalfont\AA}}
\title{Computational study of optical absorption spectra of helicenes as applied to strain sensing}

\author[1,*]{Veera Sundararaghavan}
\author[2]{Vikas Varshney}
\author[2]{Davide Simone}
\affil[1]{Department of Aerospace Engineering, University of Michigan, Ann Arbor MI 48109, U.S.A.}
\affil[2]{Materials and Manufacturing Directorate, Air Force Research Laboratory, Wright-Patterson Air Force Base, OH 45433, U.S.A.}

\affil[*]{veeras@umich.edu}


\begin{abstract}
Helicenes, a class of organic molecules consisting of ortho-fused benzene rings in a spring-like configuration have found several interesting applications in nonlinear optical materials and opto-electronic devices. Under the action of strain, i.e., via mechanical stretching or compression, the optical absorption spectra of helicenes change which can be employed for strain sensing. The present study presents a detailed investigation of the optical absorption spectra of helicenes using density functional theory along with calculations of the changes in the spectra during mechanical axial stretching or compression of helicenes. The electronic band gap followed a non-symmetric parabolic form with the amount of applied strain. A lowering of the gap in stretched or compressed helicenes compared to the pristine helicene was observed. The compressed state shows a smaller energy gap compared to tension for the same strain magnitude. A detailed inspection of the optical absorption spectra shows that compressive states show significantly lower absorption at higher optical energies (shorter wavelengths) which can provide greater sensitivity to the strain measurement. 
\end{abstract}
\begin{document}

\flushbottom
\maketitle
%
%
\thispagestyle{empty}

\section*{Introduction}

Helicenes comprise a class of organic molecules in which planar cyclic rings, such as benzene, thiophene, and their combination thereof, etc., are fused at their \textit{ortho}--positions, creating a helical structure \cite{urbano2003recent}. Even though these molecules are devoid of ‘classical’ chiral centers, they exhibit chiral characteristics that arise from their intrinsic molecular architecture (i.e., handedness of the helix), leading to their unique optical properties, such as exceptionally large optically rotatory dispersion (ORD) \cite{gingras2013one}. Due to their unique properties, helicenes and their functional derivatives have been employed in a wide variety of chemical disciplines such as catalysis \cite{narcis2014helical,yavari2014helicenes}, polymer chemistry \cite{shen2012helicenes,dhbaibi2020chiral}, biological sciences \cite{xu2004p},  non-linear optics \cite{nuckolls1998circular,wigglesworth2005chiral}, and optoelectronics \cite{yang2013circularly,fuchter20127,shi2012synthesis}.

In addition to their ability to greatly influence polarized light, helicenes present themselves as potential sensing entities due to their intrinsic spring-like characteristics, where a change in either electronic, optical, or piezoelectric response, could be examined while deforming the helicene either under compression or elongation. To date, there has been very limited non-theoretical research towards investigating the different types of responses due to external mechanical perturbations or stimuli in helicenes. Investigation of charge transport in helicenes was carried out \cite{guo2015u,vacek2015mechanical}, indicating that current along the helicene molecule depends on its length (or number of fused rings) and degree of deformation. Specifically, under stretching or compression along the helical axis, a non-monotonic (U-shaped curve) current--strain relationship was observed by Guo et al.  \cite{guo2015u}. In a different study, Vacek and co-workers computed several orders of modulation in on-off ratios in conductance,  when helicenes were utilized as a molecular switch \cite{vacek2015mechanical}.  

In the context of modulating optical features, while experimental optical absorption and circular dichroism spectra of [5]--[9] helicenes have been reported, the effect of tensile and compressive deformation of the helicenes on their optical absorption is yet to be studied. In principle, while it is possible to isolate and strain individual helicene molecules via optical tweezers \cite{neuman2008single,capitanio2013interrogating}, it is challenging to systematically investigate the effect of strain on opto-spectral response for individual helicene molecules via experiment. Theoretical investigation of such responses and identifying helicene characteristics that maximize the response sensitivity is key for their successful incorporation as \textit{in-situ}, non-invasive sensing components in various applications.

Computational chemistry provides a powerful and complementary tool for assessing the physical and chemical properties of organic molecules, often allowing direct comparison of predictions with experimental observations. In this work, we focus on modeling the electronic characteristics via density functional theory (DFT) and computation of excited states in the form of UV--Vis and electronic circular dichroism (ECD) spectra of pristine and strained helicenes using time-dependent density functional theory (TDDFT) to explore their feasibility as strain sensing entities.

\begin{figure}[h]
\centerline{\includegraphics[width=0.9\textwidth]{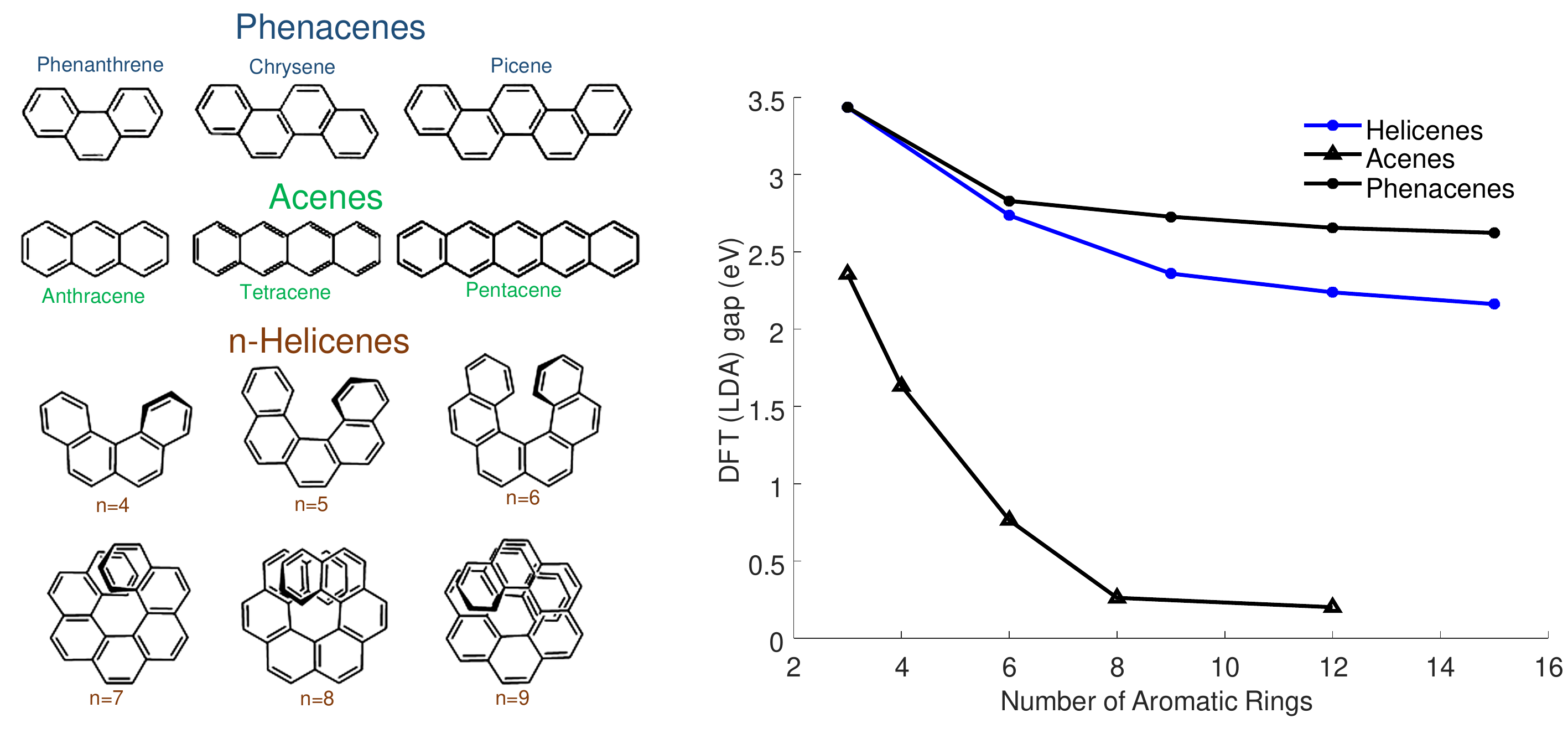}}
\caption{Comparison of the HOMO--LUMO gap of helicenes compared to its planar phenacenes and acene counterparts (shown in the left) using DFT-LDA method. The x--axis indicates the number of aromatic rings (the `n' in [n]-helicene)}
\label{1}
\end{figure}
\begin{figure}[h]
\centerline{\includegraphics[width=0.8\textwidth]{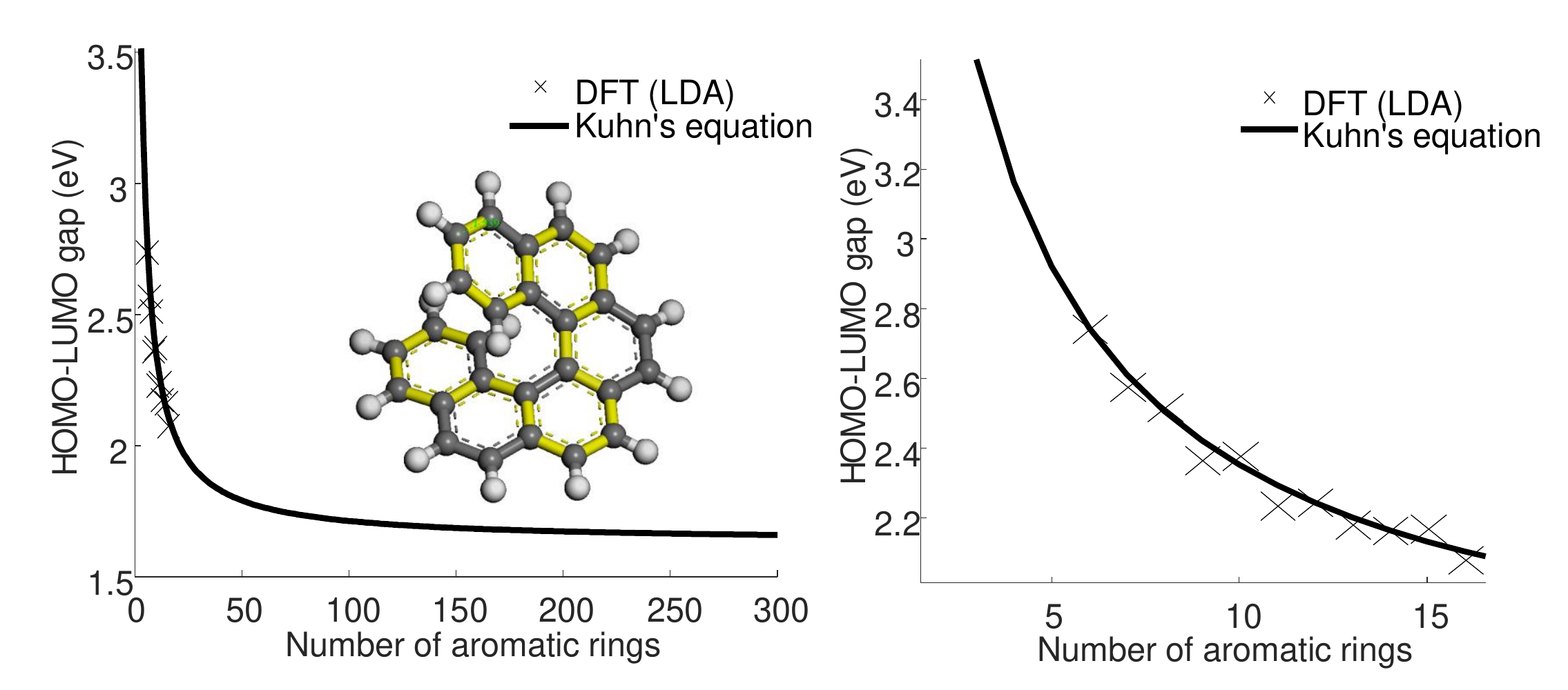}}
\centerline{(a) \hspace{3 in} (b)}
\caption{ Decrease in HOMO--LUMO gap with an increase in helicene length. (a) The trend follows Kuhn's equation (shown as inset), plotted here for even numbered helicenes. (b) Trend zoomed for shorter helicenes. Inset in (a) shows the zigzag length of even helicenes used in the formula as highlighted in yellow.}
\label{2}
\end{figure}

\section*{Results}

\subsection*{Electronic gap Prediction}
Fig. \ref{1} shows the predicted gap (from local density approximation (LDA)) between the highest occupied molecular orbital (HOMO) and lowest unoccupied molecular orbital (LUMO) for helicenes and its planar analogues (phenacenes and acenes) as a function of the number of benzene units within the modeled structures. Although LDA under-predicted the true HOMO-LUMO gap (i.e., HF predicted gap), the observed trends are in good agreement with what has been observed in experimental data as aggregated in Ref. \cite{Malloci_2011}. It should be noted that the similar HOMO-LUMO gaps of helicenes and phenacenes are due to their cross-conjugated nature of the sp$^2$ bonds in the aromatic rings relative to the conjugated structure of acenes. We find that the gap of helicenes is marginally lower than that of phenacenes and is primarily attributed to the non-planar nature of helicenes and the deviation of bond lengths associated with external aromatic bonds of the helix versus internal aromatic bonds \cite{schulman1999aromatic}.  On the contrary, no cross-conjugation occurs in acenes, leading to sharper decay in the gap with increasing acene length.  

Interestingly, the HOMO-LUMO gap for all the structures was predicted to decrease with increasing length. Due to the conjugated nature of studied structures, it is imperative that the $\pi$-electrons are involved in electronic transitions between the HOMO and LUMO. As the length of the helicene and its planar analogue increases, the conjugated network on the molecule becomes increasingly delocalized, and hence, the transition from HOMO to LUMO is expected to require progressively lower energy, as can be seen from the computed gap.

The decrease of energy gap as a function of length for conjugated organic chains is embodied in Kuhn's equation \cite{kuhn1949quantum} as follows:
\begin{equation}
    \Delta E = \frac{h^2}{8mL^2}(N+1) + V_o(1 - \frac{1}{N})
\end{equation}
where $N = 4n+2$ is the number of $\pi$ electrons, $h$ is the Planck’s constant, $m$ is the electron mass, $L$ is the zig--zag length (as described next), and $V_0$ is the HOMO--LUMO gap of an infinitely long helicene, which is a fitting parameter. The zig--zag length, $L$, of a representative even helicene is shown in yellow in the inset in Fig. \ref{2}(a) and is equal to 3$n$+3 times the C-C bond length of 1.4 {\AA} and follows the bond conjugation. Kuhn's equation is a single parameter fit based on an appropriate choice of $V_0$, the HOMO--LUMO gap of an infinitely long helicene. The value of $V_0$ is taken to be $0.06$ Ha ($1.6$ eV) in the fit shown in Fig. \ref{2}(a). 

Band gap of planar analogs have been studied in the past in the form of zero--dimensional (finite length) and one-dimensional (infinite length) graphene nano ribbons \cite{shemella2007energy}. Among these 1D ribbons, while armchair ribbons are either calculated as semiconducting or metallic depending on their width, zigzag ribbons are found to be metallic for all widths \cite{son2006energy}. 1D armchair ribbons follow a rule that the nanoribbon is metallic if the number of rows of carbon (width-wise) is 3$M$-1, where $M$ is an integer. Phenacenes and acenes analogs studied in our study can respectively be treated as finite length arm-chair (4 rows of carbon atoms) and zig-zag nanoribbons. It can be seen that acene analogs decay toward a zero HOMO-LUMO gap with increasing length, corroborating with the metallic nature of infinitely long metallic nanoribbons. Similarly, phenacenes, which have  four rows of carbon atoms along ribbon-width are expected to have a non--zero band gap at infinite lengths. The reported DFT band gap for phenacenes of infinite length is 2.6 eV\cite{shemella2007energy}. Helicenes are expected to have a lower HOMO-LUMO gap and the estimated value of ~$1.6$ eV appears reasonable. Note that Kuhn's equation does not account for features such as non--planarity of molecules other than through the chosen parameter $V_0$. Fig. \ref{2}(b) shows that odd numbered $n$--helicenes have similar energy gaps as ($n$+1) even helicenes although one would expect the gap to be slightly higher owing to the shorter zig--zag length (3$n$+2) times the C-C bond length for the same $V_0$. 

\begin{figure}[h]
\centerline{\includegraphics[width=1.1\textwidth]{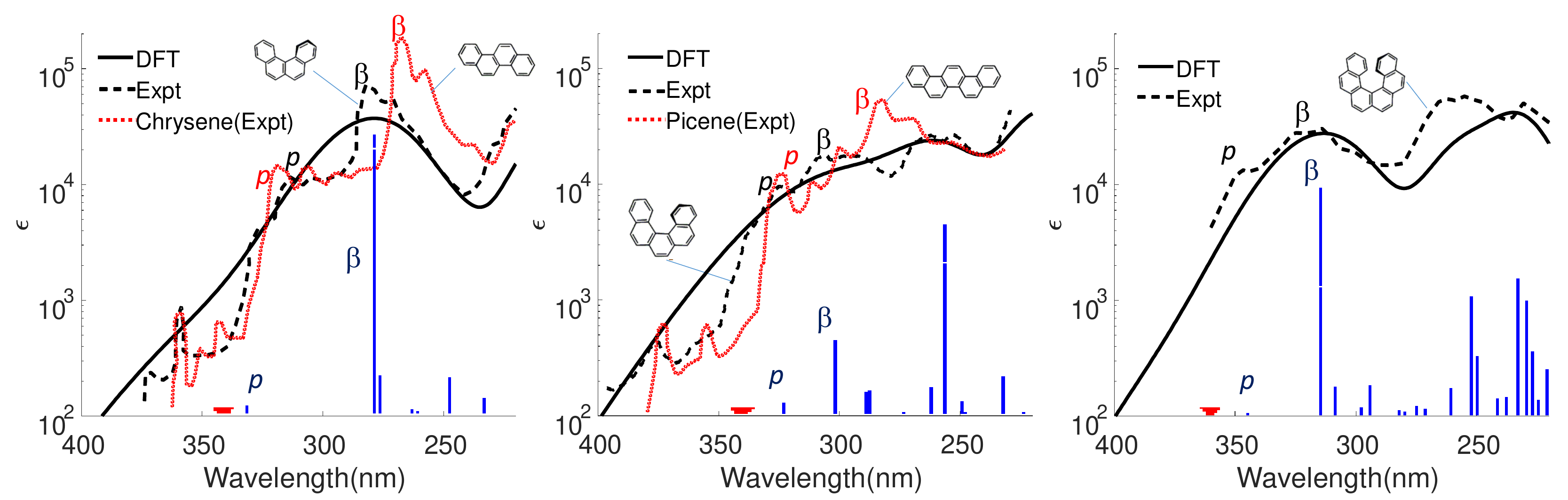}}
\centerline{(a)\hspace{2 in} (b)\hspace{2 in} (c)}
\caption{Comparison of predicted optical absorption with experimentally observed values. The bars indicate the absorption computed at the first 25 excitation frequencies. (a) [4]-helicene  (b) [5]-helicene (c) [6]-helicene. Experimental data from Ref. \cite{clar1952aromatic,newman1967optical}.}
\label{3}
\end{figure}

\subsection*{Prediction of Optical Absorption Spectra}

\begin{figure}[h]
\centerline{\includegraphics[width=1.0\textwidth]{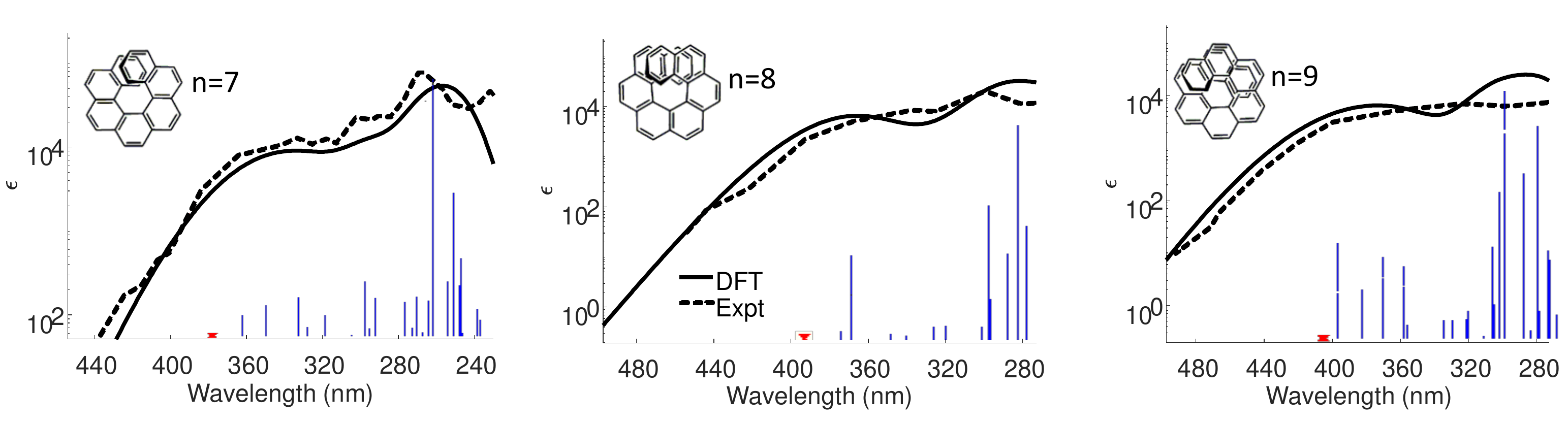}}
\centerline{(a)\hspace{2 in} (b)\hspace{2 in} (c)}
\caption{Comparison of predicted optical absorption with experimentally observed values for (a) [7]-helicene  (b) [8]-helicene (c) [9]-helicene. Experimental data from \cite{martin19681, FLAMMANGBARBIEUX1967743}}
\label{5}
\end{figure}

Next, we discuss and compare the  predicted optical absorption with the experimental literature in Figs. \ref{3}--\ref{5}. The vertical bars in these plots indicate the absorption intensity computed at the first 25 excitation frequencies and the data is directly compared against published experimental spectra for helicene molecules of various lengths reported in Refs. \cite{clar1952aromatic,newman1967optical,martin19681, FLAMMANGBARBIEUX1967743}. 

Fig(s). \ref{3}a \& \ref{3}b compare our predictions against the experimental data  of Clar and Stewart (1952) \cite{clar1952aromatic} for [4]-helicene and [5]-helicene, respectively. The figures also contain the experimental spectra of the planar phenacene counterpart (solid line) which, as previously shown in Fig. \ref{1}, shows an optical gap at a marginally lower wavelength (higher energy) compared to the helicene.  The [6]-helicene optical absorption data is compared against another experiment by Newman et al. \cite{newman1967optical} in Fig. \ref{3}(c). The optical absorption spectra of helicenes contain two prominent peaks near the HOMO--LUMO gap called the para ($p$) band and the $\beta$ band. The $p$-bands are related to fused  benzene rings, while $\beta$ peaks are associated with higher energy, high intensity B type excitations \cite{buss1996electronic,vander1968fluorescence,weigang1968low}. Both peak wavelengths are well captured by our simulations. The experimental data also shows that helicene $\beta$ peak is shifting to lower energy with respect to its phenacene analogue. This shift has been previously explained to be the result of greatly increased overlapping of H atoms due to the twisting and the subsequent loss in resonance energy \cite{clar1952aromatic}. With  [4]--, [5]-- and [6]--helicenes, coplanar overlap of end benzene rings is not possible.

Optical spectra has been analyzed in terms of electronic transitions in Table 1 below. Only the three states with the lowest energy (longest wavelengths) are analyzed. The configurations with highest contributions to the excited state are given in a notation having commonality across molecules. The HOMO-LUMO transition is always considered 1-1* regardless of system size. HOMO-1 to LUMO+1 is 2-2* etc. The percentage contribution of these orbitals to the excited state are indicated. The oscillator and rotatory strengths (in $10^{-40}$ cgs) are provided. The electronic excitations of $C_2$-symmetric helicenes can be classified according to the two
types of polarized transitions: 1-1* and 2-2* are polarized
along the short $C_2$-axis and are of the `a' type, while  1-2* and 2-1* are `b'-type excitations are polarized along the long axis (orthogonal to $C_2$). In even helicenes, the lowest energy excitations tend to be of the `b' type and for odd helicenes, these are of the `a' type, which is consistent with past calculations\cite{buss1996electronic,furche2000circular}. In the context of helicenes, the three excited states indicated correspond to $\alpha$, p and $\beta$ bands respectively. The $\alpha$ peak always consists of the low energy low intensity (`L' type) CI states. The $\beta$ bands are of the high energy high intensity (`B' type) and is prominent compared to the other two bands in helicenes of lower lengths. The prominence of the $\beta$ peak reduces with higher helicenes as described in Ref. \cite{vander1968fluorescence}, which we postulate arises from end benzene rings axially overlapping due to a complete pitch of rotation. The overlap gives rise to an enhanced probability of intra-molecular pi-stacking interactions leading to lower energy electronic transitions. 

Fig. \ref{5}(a) compares the optical gap and absorption spectra of [7]--, [8]--, and [9]-- helicenes, which involve a direct and increasing overlap of benzene rings (due to complete pitch),  with experiments reported in Refs \cite{martin19681, FLAMMANGBARBIEUX1967743}. The figure shows excellent agreement with experimental spectra and highlights the decreasing trend of lower energy optical gap with increasing helicene length. The trend is also consistent with the Kuhn's equation which predicts lower energy gaps due to extended delocalization as a function of helicene length.

\begin{figure}[h]
\centerline{\includegraphics[width=1.0\textwidth]{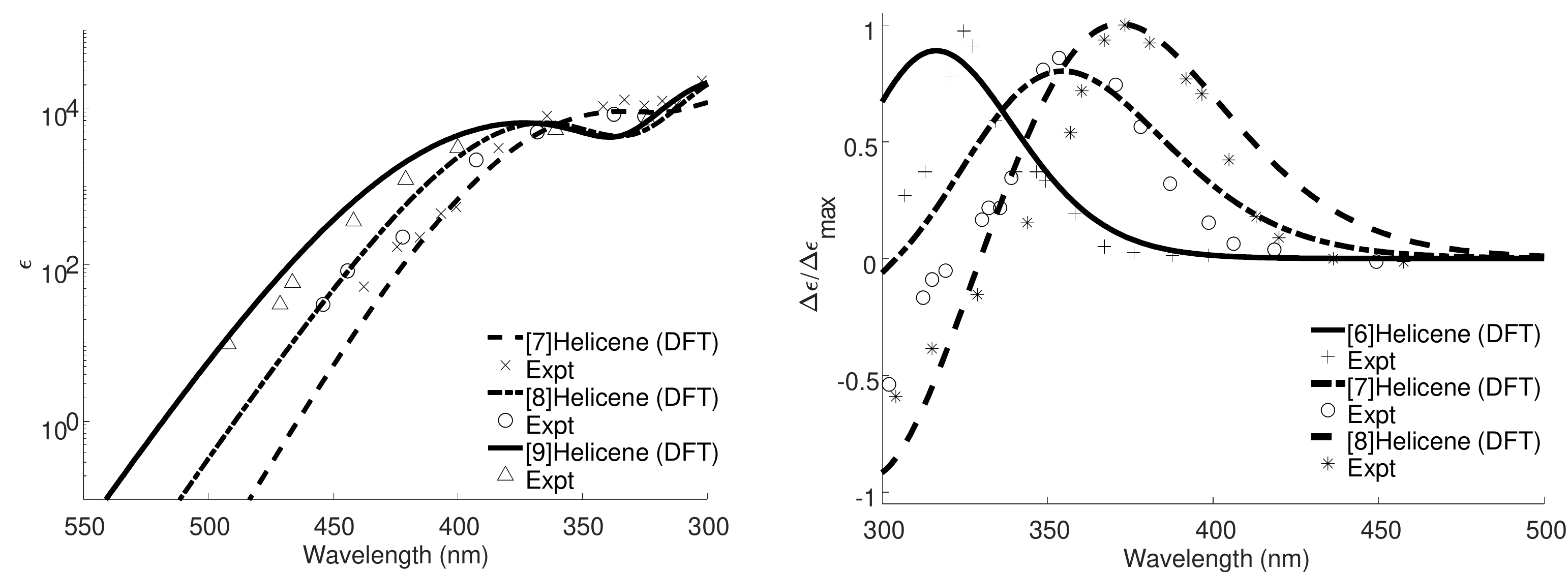}}
\centerline{(a) \hspace{3 in} (b)}
\caption{Comparison of predicted spectra against data from \cite{nakai2012theoretical,martin19681,FLAMMANGBARBIEUX1967743} (a) UV-Vis spectra (b) ECD spectra.}
\label{6}
\end{figure}

\begin{figure}[h]
\centerline{\includegraphics[width=0.9\textwidth]{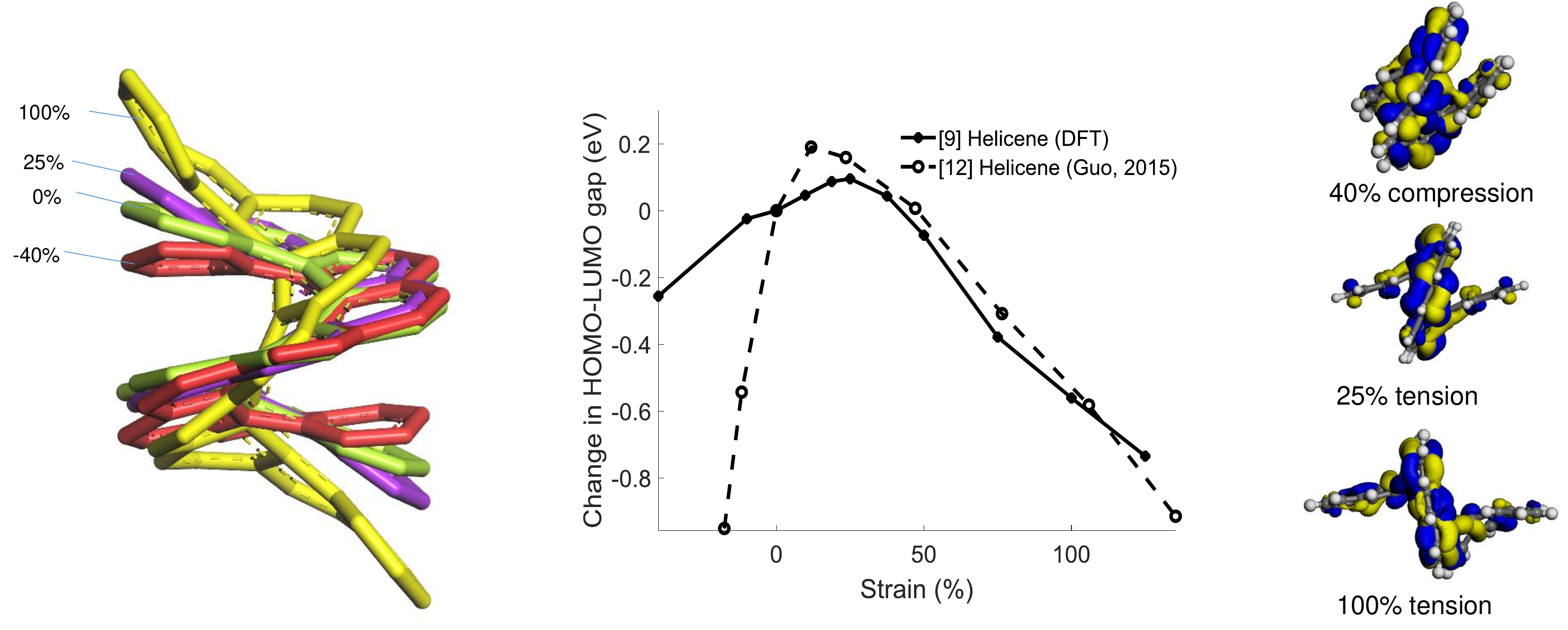}}
\centerline{(a) \hspace{2.0 in} (b)\hspace{2.2 in} (c)}
\caption{(a) Axial straining of [9]--helicene. (b) The change in HOMO-LUMO gap as a function of strain for [9]-helicene, the data is plotted against data from Guo et al (2015) for [12]-helicene. (c) The HOMO and LUMO orbitals (plotted together) for 40\% compression, 25\% and 100\% tension.}
\label{7}
\end{figure}

The optical absorption spectra for [7]--, [8]--, and [9]-- helicenes are aggregated together along with experimental data from Refs \cite{martin19681, FLAMMANGBARBIEUX1967743} in Fig. \ref{6}(a). In addition to optical absorption spectra, we also calculated the ECD spectra for [6]-- to [8]--helicene and compare the findings against experimental data in Ref. \cite{nakai2012theoretical}. The predicted as well as experimental ECD spectra are shown in Fig. \ref{6}(b). We find that the predicted ECD spectra compare favorably with experiments. It should be noted that due to the calculation of only 25 excited states, the higher wavelength (low energy) results can be more accurately compared against the experiments. Helicenes with even-numbered rings typically show a higher peak differential absorption of left and right circularly polarized light as compared to odd--helicenes as seen in both experiments and DFT simulations. The good agreement between measured and predicted optical spectra (as discussed in Figs 3 and 4) validated the level of accuracy/fidelity of first-principle calculations necessary to carry out helicene deformation (compression and elongation) calculations, which we discuss next. 

\begin{table}[h!]
\centering
\begin{tabular}{||c c c c c c||} 
 \hline
 [n] & \%contribution & CI parentage & Wavelength (nm) & Oscillator strength & Rotatory strength \\ [0.5ex] 
 \hline\hline
4&	52,47&	2-1*,1-2*&	339.5&	0&	-0.1\\
&	88,10&	2-2*,1-1*&	330.59&	0.02&	-19\\
&	48,45&	1-2*,2-1*&	279.82&	0.78&	112\\&&&&&\\
	
5&	40,22&	1-1*,2-2*&	342.13&	0&	-3\\
&	53,32&	2-1*,1-1*&326.34&	0.03&	88\\
&	49,20&	1-2*,2-2*&	305.19&	0.17&	197\\&&&&&\\
	
6&	57,40&	2-1*,1-2*&	358.35&	0&	2\\
&	22,77&	2-2*,1-1*&	343.64&	0&	-0.6\\
&	55,39&	1-2*,2-1*&	315.59&	0.42&	686\\&&&&&\\

7&	62,36&	1-1*,2-2*&	377.41&	0&	3\\
&	95,2.7&	2-1*,1-2*&	361.79&	0.03&	274\\
&78,17&	1-2*,3-1*&	349.44&	0.05&	291\\&&&&&\\

8&	60,28&	1-2*,2-1*&	392.47&	0&	25\\
&	62,21&	1-1*,2-2*&	373.21&	0.01&	112\\
&	66,21&	2-1*,1-2*&	368.05&	0.08&	702\\ [1ex] 
 \hline
\end{tabular}
\caption{Calculated spectral data (\%contribution corresponding to CI parentage, corresponding wavelength, oscillator and rotatory strengths (in $10^{-40}$ cgs) are indicated. Only top three long wavelength transitions are considered.}
\label{table:1}
\end{table}

\subsection*{Effect of Applied Strain on HOMO--LUMO gap \& Spectra Modulation}
In this section, we discuss how the tensile and compressive axial deformation affects the optical absorption spectra of helicene molecules towards exploring their potential as embedded sensing modalities using the representative example of [9]--helicene. In this context, we start our discussion with the visualization of different strained states of [9]--helicene as well as the estimation of change in respective HOMO--LUMO gap with respect to an undeformed helicene as shown in Fig(s). \ref{7}(a) \& (b). It can be seen from the figure that the change in gap follows a non-monotonic, asymmetric, inverted parabola-like form with respect to the amount of applied strain. For both large compressive and tensile deformation, a lower band gap is observed as compared to the pristine, un-deformed helicene. The parabolic form is unique to helical graphene structure and its analogs. For example, under bending as shown in Ref \cite{farajian2003electronic}, bandgap of semiconducting nanotubes monotonically decreases with bending, while that of metallic nanotubes would increase with bending.  In helicenes, the LUMO energies are  sensitive to straining and follow a similar parabolic form to strain as the HOMO--LUMO gap itself. This unique trend in LUMO energies can be associated with the twisted structure of helicenes. On compression, the gap decreases due to the higher overlap of orbitals and subsequent loss of resonance energy. Fig. \ref{7}(c)(top) depicts the HOMO and LUMO orbitals for helicene axially strained under compression (40\% strain) depicting overlap of orbitals from the center towards the benzene units at the free ends. 

The behavior on stretching is more interesting, as one would expect the gap to increase towards those what is observed in planar phenacenes. This is indeed the case until a critical strain of 25\%. Beyond this strain, although HOMO energies continue to increase upon tensile straining as expected, the LUMO energies show an opposing trend, i.e., decrease upon further tensile straining. In a previous study by Guo et al. \cite{guo2015u} on [12]-helicene, a critical strain of 23.53\% was reported where the inflection occurs towards a decreasing gap under increased tension (curve depicted in Fig. \ref{7}(b)) . 

\begin{figure}[h]
\centerline{\includegraphics[width=0.6\textwidth]{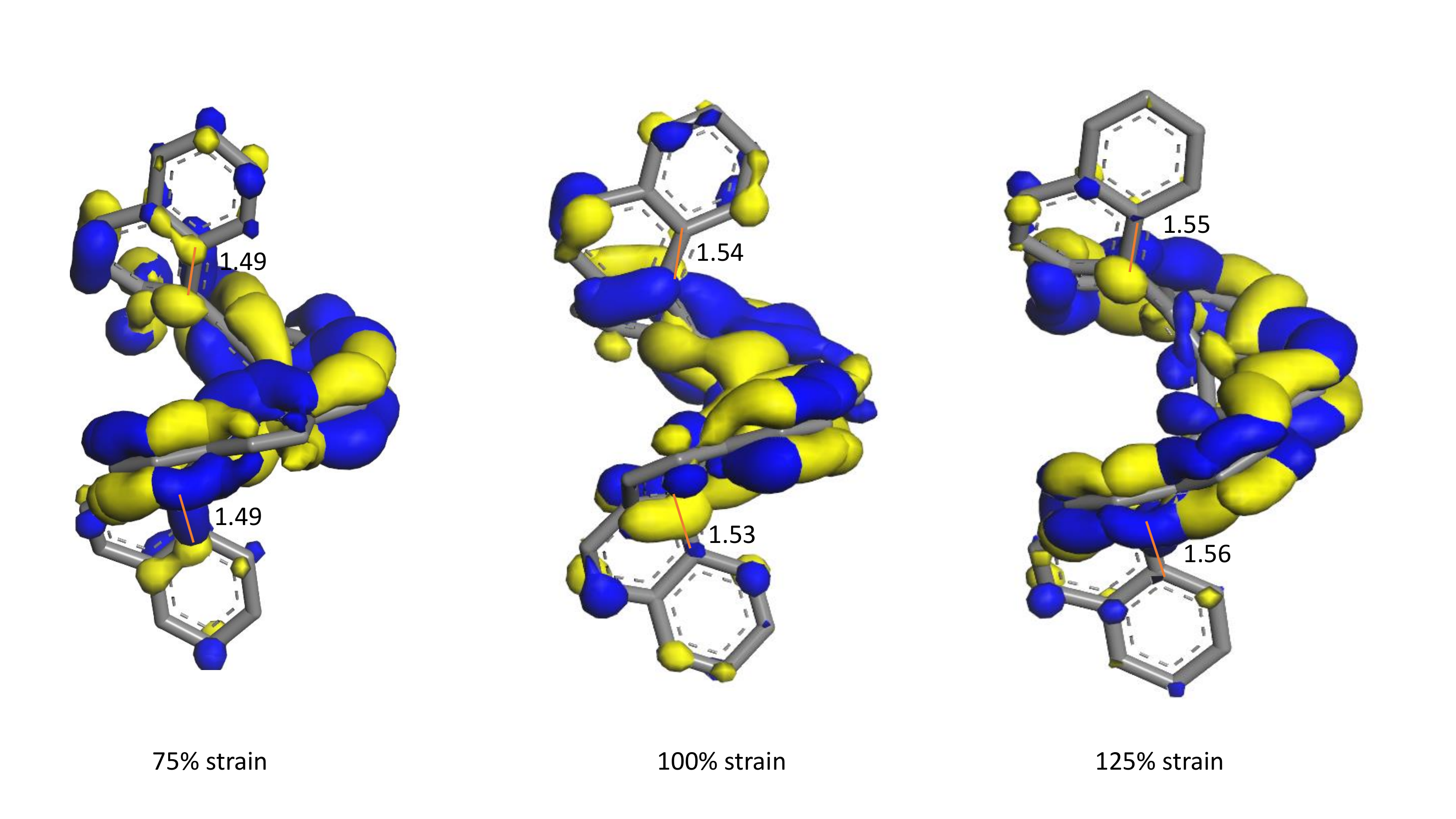}}
\caption{HOMO and LUMO orbitals (together) are indicated at three strain levels around failure strain.}
\label{s2}
\end{figure}

In Fig. \ref{7}(c), we show the behavior of orbital wave functions at the critical strain of 25\% where the highest gap is realized. At this strain, overlap of orbitals out-of-plane of the benzene rings as seen at 40\% compression is completely removed, leading to an increase in the HOMO-LUMO gap. Beyond this critical strain, the HOMO-LUMO gap begins to decrease. The HOMO-LUMO orbital structure is illustrated for the case of 100\% tensile strain in Fig. \ref{7}(c)(bottom). In Ref. \cite{guo2015u}, it was found that stretching beyond 23.53\% tensile strain tends to localize the LUMO wave function toward the inner  helix. However, that study enforced proportional elongation of all helicene atoms before optimization which could have biased the structure towards a geometry with a homogeneous strain distribution. In our study, we see that the geometry optimization after straining leads to inhomogeneous straining of helicenes (see Fig \ref{7}(a)). At 100\% strain, the free ends take up significant deformation via bending of the chain, while the  central helix is subject to a relatively smaller amount of stretching. Alternating inner helix bonds are highly deformed during tension, with a few bonds stretching towards the $sp^3$ bond failure length of 1.54 \angstrom (at 100\% strain) making conjugation along the inner helix unlikely. Fig \ref{s2} shows the inner helix bond lengths at 100\% strain at which the bond failure length close to 1.54 \angstrom~is reached. Bond lengths at 75\% and 125\% strains are also shown as comparison (also see Fig S1 in supplementary for additional bond lengths along inner helix). 

\begin{figure}[h]
\centerline{\includegraphics[width=0.6\textwidth]{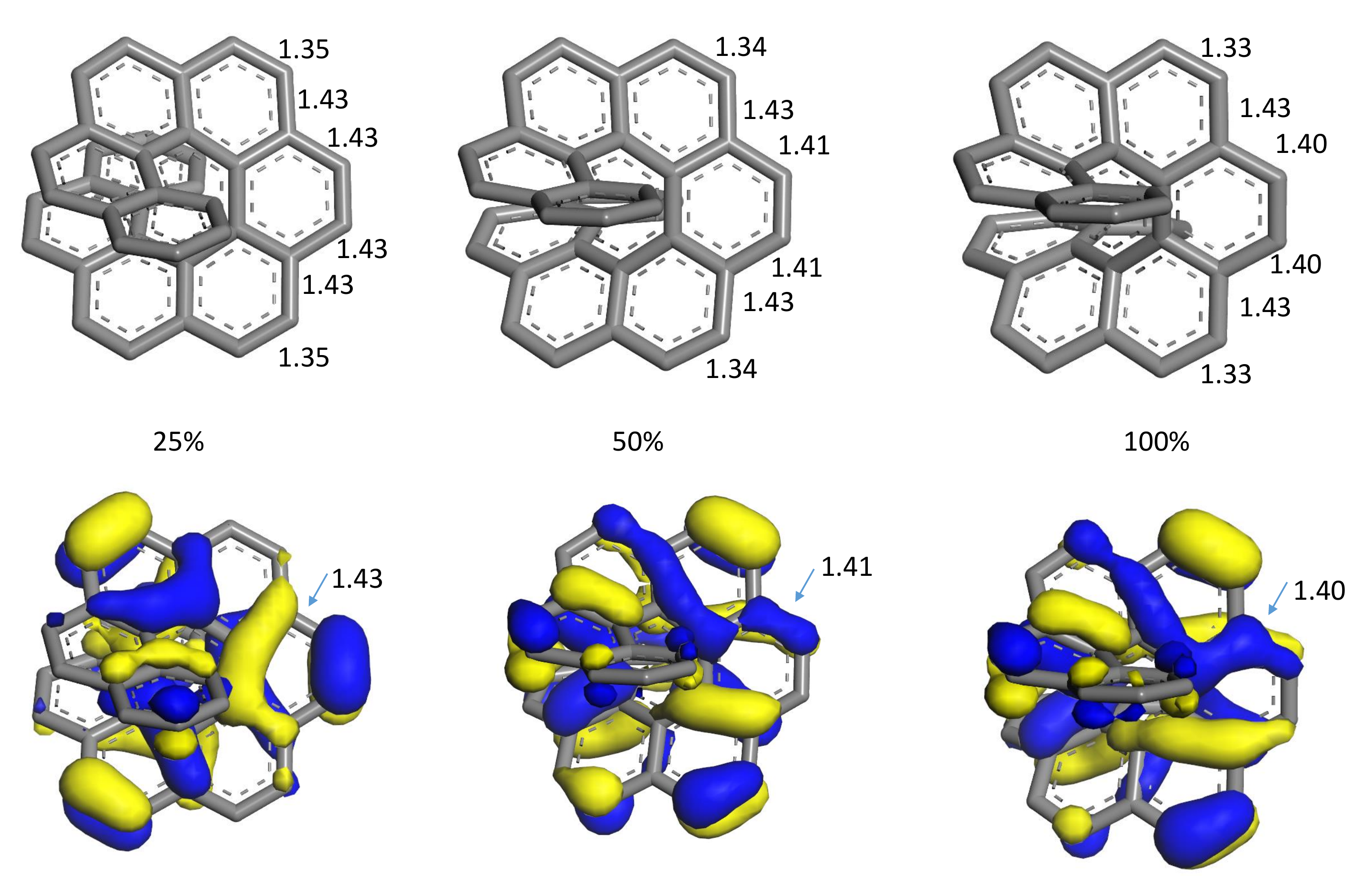}}
\caption{Change of few outer helix bond lengths towards 1.4 \angstrom, the $sp^2$ benzene bond length, at higher strains tends to improve conjugation along these bonds. The bonds indicated with arrows. Strain levels indicated below each case.}
\label{s3}
\end{figure}

A different explanation of lower gap upon elongation compared to Ref. \cite{guo2015u} can be provided in terms of the outer helix bond lengths in strained helicenes. An $sp^2$ carbon in an alkene is 1.34 \angstrom~in length, an $sp^3$ carbon is 1.46 \angstrom~in length; an $sp^2$ benzene bond is 1.40 \angstrom~in length. The inner helix of the helicene has bond lengths of 1.45 \angstrom~similar to $sp^3$ carbon and is thus not very aromatic or conjugated. The outer helix has alternating bond lengths of 1.36 \angstrom~and 1.43 \angstrom~and are more conjugated and aromatic than the inner helix \cite{schulman1999aromatic}. A decrease in the HOMO-LUMO gap implies better conjugation of bonds. However, one would expect bond lengths to increase during tensile straining greater than 25\% lowering the strength of bonds and affecting conjugation in the structure. On inspection of the structures, it is seen that length of longer bonds in the outer helix decreases towards 1.40 \angstrom~with increasing strain. Figure \ref{s3} shows the decrease of certain outer helix bond lengths, indicated with arrows, towards 1.4 \angstrom~with straining and the accompanying change in the distribution of the highest occupied orbital wavefunctions. Fig. \ref{7}(c)(bottom) depicts the HOMO and LUMO orbitals for helicene under tension (at 100\% strain) showing that orbitals continue to localize with improved conjugation along the outer helix compared to the 25\% strain case. Equalization of these outer helix bond lengths towards the $sp^2$ benzene bond length contributes to lower band gap through improved conjugation in the outer helix. 

Additional analysis of the change in electronic transitions upon straining has been included in  Table 2. Although the change in the two lowest energy excitations is small across the large change in strains, the $\beta$ peak (third entry) shows a larger sensitivity, with a red shift seen under compression and a blue shift under tension. This can be attributed to the improved out-of-plane conjugation observed under compression, lowering the excitation energy. The strength of the $\beta$ peak is also seen to significantly increase under tension allowing differentiation of the strain state.

\begin{table}[h!]
\centering
\begin{tabular}{||c c c c c c||} 
 \hline
 Strain & \%contribution & CI parentage & Wavelength (nm) & Oscillator strength & Rotatory strength \\ [0.5ex] 
 \hline\hline
0&	57,30&	1-1*,2-2*&	404.4&	0&	-0.4\\
&	88,8&	2-1*,1-2*&	395.55&	0.04&	398\\
&	60,33&	1-2*,3-1*&	381.2&	0.02&	271\\&&&&&\\
	
-40&	74,20&	1-1*,3-2*&	404.82&	0&	-20\\
&	83,13&	2-1*,1-2*&	403.64&	0&	22\\
&	92,2&	3-1*,1-2*&	398.4&	0.03&	358\\&&&&&\\
	
50&	54,37&	1-1*,2-2*&	407.97&	0&	-0.01\\
&	66,27&	2-1*,1-2*&	387.7&	0.03&	116\\
&	64,27&	1-2*,2-1*&	369.72&	0.3& 1197\\[1ex] 
 \hline
\end{tabular}
\caption{Calculated spectral data for [9]-helicene as a function of strain (other entries similar to table 1). Only the top two main contributing configurations are indicated.}
\label{table:1}
\end{table}

\section*{Discussion}

The spring-like configuration of Helicene molecules leads to a smaller HOMO-LUMO gap than that of its planar phenacene analogs. As the length of the helicene chain increases, the conjugated network on the molecule becomes more spread out, and hence, the transition from HOMO to LUMO requires progressively lower energy, following Kuhn's equation. The computed optical absorption spectra of selected helicene molecules are found to  compare well against published experimental data. The spectra contain two prominent peaks near the HOMO--LUMO gap called the para ($p$) band and the $\beta$ band. The helicene peaks shift to lower energy with respect to its phenacene analogue as a result of increased overlapping of atoms due to the twisting.

\begin{figure}[h]
\centerline{\includegraphics[width=0.6\textwidth]{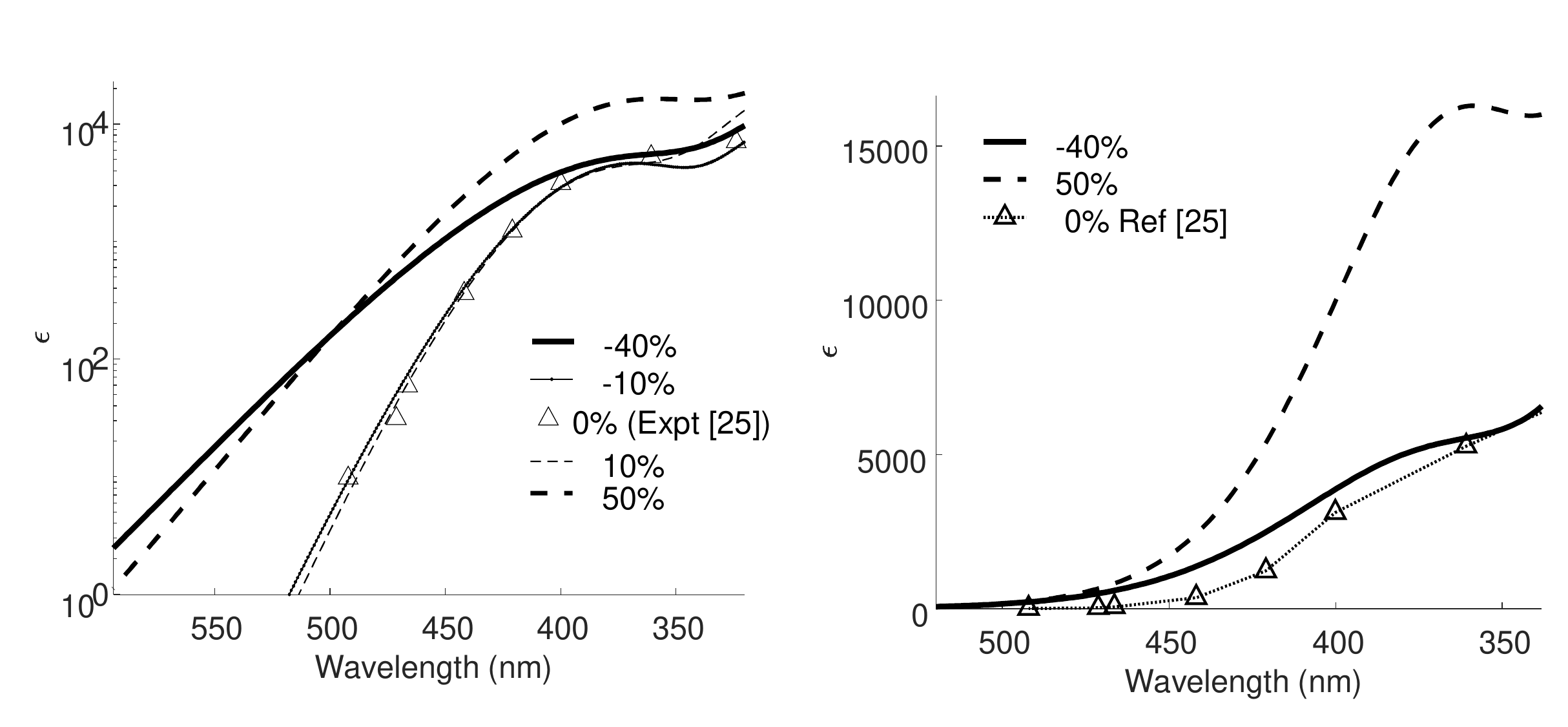}}
\centerline{(a)\hspace{3.5 in} (b)}
\caption{(a) Optical absorption spectra for selected strain states (a) comparison at wavelengths near the optical gap (b) Comparison over a larger set of wavelengths for 50 \% tension and 40\% compression.}
\label{8}
\end{figure}

\begin{figure}[h]
\centerline{\includegraphics[width=0.6\textwidth]{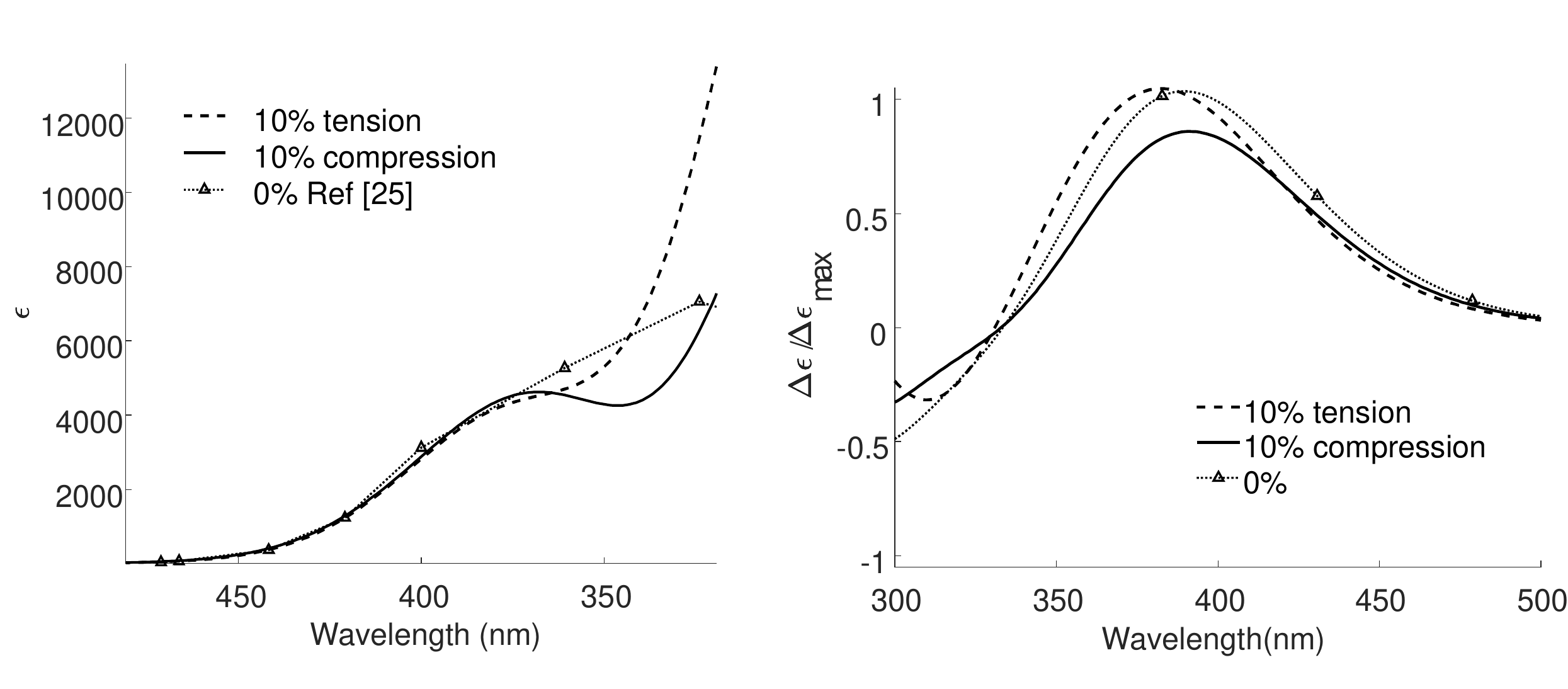}}
\centerline{(a)\hspace{3.5 in} (b)}
\caption{(a) Optical absorption compared for 10\% compression or tension states (b) Comparison of the ECD spectra for these strain states.}
\label{9}
\end{figure}

When helicenes are strained, the changes in the HOMO-LUMO gap followed a non-symmetric parabolic trend with the amount of straining.  Compressed helicenes have a lower gap due to increased conjugation along the helical axis due to chain overlap. The inflection in the HOMO--LUMO gap with tensile strain can be explained as a transition from increasing HOMO-LUMO gap due to decreased overlap to decreasing HOMO-LUMO gap due to improved conjugation as few outer helix bonds reach the $sp^2$ benzene bond length during stretching. The observed parabolic trend poses an issue for strain sensing as the fact that for the same measured optical gap, there are two possible strain states of opposite manner (either compression or elongation). This can be seen in the optical absorption spectra plotted for different strain states in Fig. \ref{8}(a) where selected compressive and tensile strain states show similar spectra near the optical gap. However, over a larger range of wavelengths, compressive states show lower absorption at higher optical energies (lower wavelengths), resolving the issue of differentiating the sign of strains. Fig. \ref{8}(b) and Fig \ref{9}(a) compare the absorption spectra (linear scale) for tension versus compression states. It is seen that the para ($p$-band) and $\beta$ peaks are decreased due to the overlap of orbitals in compression states. Similar behavior is also seen in the ECD spectra (Fig. \ref{9}b). Such a feature would allow for a higher sensitivity measurement of strain states, coupled with optical gap measurements. 

These insights provide promise for helicenes as a potential strain sensing molecule when used in composite materials. Additional investigations on  how the optical gap as well as optical spectra gets modulated with respect to different helicene parameters such as helicene length, heteroatom functionalization (S, N, P atoms), ring substituents, and electronic doping, would be valuable in highlighting their capability for future strain sensing applications.

\section*{Methods}

All structures were generated and optimized using COMPASS force-field in Forcite code \cite{accelrys2010materials} to be used as starting structures for first-principle calculations. Preliminary band gap calculations for various helicenes (as a right handed helix) as well as their planar phenacene and acene analogs (see Fig. \ref{1}) were carried out using DFT calculations with local density approximation with Vosko-Wilk-Nusair\cite{vwn}(LDA-VWN) exchange correlation. In order to generate optimized geometries of various helicene structures for TDDFT calculations,
Hartree-Fock (HF) theory was employed as it provides exact exchange correlation. Furche et al (2000) \cite{furche2000circular} presented the first TDDFT simulations of circular dichroism spectra of [4]-[7] carbohelicenes using geometry relaxed using Hatree-Fock theory. They found that DFT calculations with GGA and hybrid functionals lead to inferior results through comparison with experimental molecular structure. Double numeric basis sets with polarization functions were used with an energy convergence criteria of 1.0e$^{-4}$ Ha to obtain the optimized structures of different  helicenes studied here. 

Optical absorption simulations of helicenes were carried out using TDDFT method in Gaussian-09 code (frequency space implementation is described in detail in Ref. \cite{stratmann1998efficient}). The first 25 excitation frequencies were computed to model the optical spectra. The B3LYP exchange correlation functional has been used based on its past success in modeling polyaromatic hydrocarbons\cite{cardia2014effects,cardia2016electronic,dardenne2017tuning}. In the context of carbohelienes, Nakai (2012) \cite{nakai2012theoretical} performed a comparison of different exchange functionals and found that the amount of exact exchange in the DFT functionals plays an important role. B3LYP was found to reasonably reproduce the experimental CD spectrum of [6] helicene, although slightly underestimating the experimental CD intensities. The  correlation consistent polarized valence double zeta (ccPVDZ) basis set was chosen, which has been successfully employed in the past for polyaromatic hydrocarbons \cite{martin1996vibrational} including carbohelicenes\cite{schaack2019helicene} in combination with B3LYP exchange correlation functional. The smoothed response was computed using a peak broadening parameter (half width of the peak at half height) of 0.233 eV. To model the effect of helicene deformation on optical properties, the helicenes were strained along the helical axis direction by displacing the  end aromatic carbon atoms of the helical structure by a displacement corresponding to the needed strain level (in both elongation and compression). Then, the structure was re-optimized using HF theory (as detailed above) but with the two end-atoms fixed. The optical absorption spectra of the strained structure were then computed using TDDFT.

\bibliography{sample}

\begin{thebibliography}{10}
\urlstyle{rm}
\expandafter\ifx\csname url\endcsname\relax
  \def\url#1{\texttt{#1}}\fi
\expandafter\ifx\csname urlprefix\endcsname\relax\def\urlprefix{URL }\fi
\expandafter\ifx\csname doiprefix\endcsname\relax\def\doiprefix{DOI: }\fi
\providecommand{\bibinfo}[2]{#2}
\providecommand{\eprint}[2][]{\url{#2}}

\bibitem{urbano2003recent}
\bibinfo{author}{Urbano, A.}
\newblock \bibinfo{journal}{\bibinfo{title}{Recent developments in the
  synthesis of helicene-like molecules}}.
\newblock {\emph{\JournalTitle{Angewandte Chemie International Edition}}}
  \textbf{\bibinfo{volume}{42}}, \bibinfo{pages}{3986--3989}
  (\bibinfo{year}{2003}).

\bibitem{gingras2013one}
\bibinfo{author}{Gingras, M.}
\newblock \bibinfo{journal}{\bibinfo{title}{One hundred years of helicene
  chemistry. part 3: applications and properties of carbohelicenes}}.
\newblock {\emph{\JournalTitle{Chemical Society Reviews}}}
  \textbf{\bibinfo{volume}{42}}, \bibinfo{pages}{1051--1095}
  (\bibinfo{year}{2013}).

\bibitem{narcis2014helical}
\bibinfo{author}{Narcis, M.~J.} \& \bibinfo{author}{Takenaka, N.}
\newblock \bibinfo{journal}{\bibinfo{title}{Helical-chiral small molecules in
  asymmetric catalysis}}.
\newblock {\emph{\JournalTitle{European Journal of Organic Chemistry}}}
  \textbf{\bibinfo{volume}{2014}}, \bibinfo{pages}{21--34}
  (\bibinfo{year}{2014}).

\bibitem{yavari2014helicenes}
\bibinfo{author}{Yavari, K.} \emph{et~al.}
\newblock \bibinfo{journal}{\bibinfo{title}{Helicenes with embedded phosphole
  units in enantioselective gold catalysis}}.
\newblock {\emph{\JournalTitle{Angewandte Chemie}}}
  \textbf{\bibinfo{volume}{126}}, \bibinfo{pages}{880--884}
  (\bibinfo{year}{2014}).

\bibitem{shen2012helicenes}
\bibinfo{author}{Shen, Y.} \& \bibinfo{author}{Chen, C.-F.}
\newblock \bibinfo{journal}{\bibinfo{title}{Helicenes: synthesis and
  applications}}.
\newblock {\emph{\JournalTitle{Chemical reviews}}}
  \textbf{\bibinfo{volume}{112}}, \bibinfo{pages}{1463--1535}
  (\bibinfo{year}{2012}).

\bibitem{dhbaibi2020chiral}
\bibinfo{author}{Dhbaibi, K.} \emph{et~al.}
\newblock \bibinfo{journal}{\bibinfo{title}{Chiral
  diketopyrrolopyrrole-helicene polymer with efficient red circularly polarized
  luminescence}}.
\newblock {\emph{\JournalTitle{Frontiers in chemistry}}}
  \textbf{\bibinfo{volume}{8}}, \bibinfo{pages}{237} (\bibinfo{year}{2020}).

\bibitem{xu2004p}
\bibinfo{author}{Xu, Y.} \emph{et~al.}
\newblock \bibinfo{journal}{\bibinfo{title}{(p)-helicene displays chiral
  selection in binding to z-dna}}.
\newblock {\emph{\JournalTitle{Journal of the American Chemical Society}}}
  \textbf{\bibinfo{volume}{126}}, \bibinfo{pages}{6566--6567}
  (\bibinfo{year}{2004}).

\bibitem{nuckolls1998circular}
\bibinfo{author}{Nuckolls, C.} \emph{et~al.}
\newblock \bibinfo{journal}{\bibinfo{title}{Circular dichroism and uv- visible
  absorption spectra of the langmuir- blodgett films of an aggregating
  helicene}}.
\newblock {\emph{\JournalTitle{Journal of the American Chemical Society}}}
  \textbf{\bibinfo{volume}{120}}, \bibinfo{pages}{8656--8660}
  (\bibinfo{year}{1998}).

\bibitem{wigglesworth2005chiral}
\bibinfo{author}{Wigglesworth, T.~J.}, \bibinfo{author}{Sud, D.},
  \bibinfo{author}{Norsten, T.~B.}, \bibinfo{author}{Lekhi, V.~S.} \&
  \bibinfo{author}{Branda, N.~R.}
\newblock \bibinfo{journal}{\bibinfo{title}{Chiral discrimination in
  photochromic helicenes}}.
\newblock {\emph{\JournalTitle{Journal of the American Chemical Society}}}
  \textbf{\bibinfo{volume}{127}}, \bibinfo{pages}{7272--7273}
  (\bibinfo{year}{2005}).

\bibitem{yang2013circularly}
\bibinfo{author}{Yang, Y.}, \bibinfo{author}{Da~Costa, R.~C.},
  \bibinfo{author}{Fuchter, M.~J.} \& \bibinfo{author}{Campbell, A.~J.}
\newblock \bibinfo{journal}{\bibinfo{title}{Circularly polarized light
  detection by a chiral organic semiconductor transistor}}.
\newblock {\emph{\JournalTitle{Nature Photonics}}}
  \textbf{\bibinfo{volume}{7}}, \bibinfo{pages}{634--638}
  (\bibinfo{year}{2013}).

\bibitem{fuchter20127}
\bibinfo{author}{Fuchter, M.~J.} \emph{et~al.}
\newblock \bibinfo{journal}{\bibinfo{title}{[7]-helicene: a chiral molecular
  tweezer for silver (i) salts}}.
\newblock {\emph{\JournalTitle{Dalton Transactions}}}
  \textbf{\bibinfo{volume}{41}}, \bibinfo{pages}{8238--8241}
  (\bibinfo{year}{2012}).

\bibitem{shi2012synthesis}
\bibinfo{author}{Shi, L.} \emph{et~al.}
\newblock \bibinfo{journal}{\bibinfo{title}{Synthesis, structure, properties,
  and application of a carbazole-based diaza [7] helicene in a
  deep-blue-emitting oled}}.
\newblock {\emph{\JournalTitle{Chemistry--A European Journal}}}
  \textbf{\bibinfo{volume}{18}}, \bibinfo{pages}{8092--8099}
  (\bibinfo{year}{2012}).

\bibitem{guo2015u}
\bibinfo{author}{Guo, Y.-D.}, \bibinfo{author}{Yan, X.-H.},
  \bibinfo{author}{Xiao, Y.} \& \bibinfo{author}{Liu, C.-S.}
\newblock \bibinfo{journal}{\bibinfo{title}{U-shaped relationship between
  current and pitch in helicene molecules}}.
\newblock {\emph{\JournalTitle{Scientific reports}}}
  \textbf{\bibinfo{volume}{5}}, \bibinfo{pages}{16731} (\bibinfo{year}{2015}).

\bibitem{vacek2015mechanical}
\bibinfo{author}{Vacek, J.}, \bibinfo{author}{Chocholou{\v{s}}ov{\'a}, J.~V.},
  \bibinfo{author}{Star{\'a}, I.~G.}, \bibinfo{author}{Star{\`y}, I.} \&
  \bibinfo{author}{Dubi, Y.}
\newblock \bibinfo{journal}{\bibinfo{title}{Mechanical tuning of conductance
  and thermopower in helicene molecular junctions}}.
\newblock {\emph{\JournalTitle{Nanoscale}}} \textbf{\bibinfo{volume}{7}},
  \bibinfo{pages}{8793--8802} (\bibinfo{year}{2015}).

\bibitem{neuman2008single}
\bibinfo{author}{Neuman, K.~C.} \& \bibinfo{author}{Nagy, A.}
\newblock \bibinfo{journal}{\bibinfo{title}{Single-molecule force spectroscopy:
  optical tweezers, magnetic tweezers and atomic force microscopy}}.
\newblock {\emph{\JournalTitle{Nature methods}}} \textbf{\bibinfo{volume}{5}},
  \bibinfo{pages}{491--505} (\bibinfo{year}{2008}).

\bibitem{capitanio2013interrogating}
\bibinfo{author}{Capitanio, M.} \& \bibinfo{author}{Pavone, F.~S.}
\newblock \bibinfo{journal}{\bibinfo{title}{Interrogating biology with force:
  single molecule high-resolution measurements with optical tweezers}}.
\newblock {\emph{\JournalTitle{Biophysical journal}}}
  \textbf{\bibinfo{volume}{105}}, \bibinfo{pages}{1293--1303}
  (\bibinfo{year}{2013}).

\bibitem{Malloci_2011}
\bibinfo{author}{Malloci, G.}, \bibinfo{author}{Cappellini, G.},
  \bibinfo{author}{Mulas, G.} \& \bibinfo{author}{Mattoni, A.}
\newblock \bibinfo{journal}{\bibinfo{title}{Electronic and optical properties
  of families of polycyclic aromatic hydrocarbons: A systematic
  (time-dependent) density functional theory study}}.
\newblock {\emph{\JournalTitle{Chemical Physics}}}
  \textbf{\bibinfo{volume}{384}}, \bibinfo{pages}{19–27},
  \doiprefix\url{10.1016/j.chemphys.2011.04.013} (\bibinfo{year}{2011}).

\bibitem{schulman1999aromatic}
\bibinfo{author}{Schulman, J.~M.} \& \bibinfo{author}{Disch, R.~L.}
\newblock \bibinfo{journal}{\bibinfo{title}{Aromatic character of [n] helicenes
  and [n] phenacenes}}.
\newblock {\emph{\JournalTitle{The Journal of Physical Chemistry A}}}
  \textbf{\bibinfo{volume}{103}}, \bibinfo{pages}{6669--6672}
  (\bibinfo{year}{1999}).

\bibitem{kuhn1949quantum}
\bibinfo{author}{Kuhn, H.}
\newblock \bibinfo{journal}{\bibinfo{title}{A quantum-mechanical theory of
  light absorption of organic dyes and similar compounds}}.
\newblock {\emph{\JournalTitle{The Journal of chemical physics}}}
  \textbf{\bibinfo{volume}{17}}, \bibinfo{pages}{1198--1212}
  (\bibinfo{year}{1949}).

\bibitem{shemella2007energy}
\bibinfo{author}{Shemella, P.}, \bibinfo{author}{Zhang, Y.},
  \bibinfo{author}{Mailman, M.}, \bibinfo{author}{Ajayan, P.~M.} \&
  \bibinfo{author}{Nayak, S.~K.}
\newblock \bibinfo{journal}{\bibinfo{title}{Energy gaps in zero-dimensional
  graphene nanoribbons}}.
\newblock {\emph{\JournalTitle{Applied Physics Letters}}}
  \textbf{\bibinfo{volume}{91}}, \bibinfo{pages}{042101}
  (\bibinfo{year}{2007}).

\bibitem{son2006energy}
\bibinfo{author}{Son, Y.-W.}, \bibinfo{author}{Cohen, M.~L.} \&
  \bibinfo{author}{Louie, S.~G.}
\newblock \bibinfo{journal}{\bibinfo{title}{Energy gaps in graphene
  nanoribbons}}.
\newblock {\emph{\JournalTitle{Physical review letters}}}
  \textbf{\bibinfo{volume}{97}}, \bibinfo{pages}{216803}
  (\bibinfo{year}{2006}).

\bibitem{clar1952aromatic}
\bibinfo{author}{Clar, E.} \& \bibinfo{author}{Stewart, D.}
\newblock \bibinfo{journal}{\bibinfo{title}{Aromatic hydrocarbons. lxiii.
  resonance restriction and the absorption spectra of aromatic hydrocarbons1}}.
\newblock {\emph{\JournalTitle{Journal of the American Chemical Society}}}
  \textbf{\bibinfo{volume}{74}}, \bibinfo{pages}{6235--6238}
  (\bibinfo{year}{1952}).

\bibitem{newman1967optical}
\bibinfo{author}{Newman, M.~S.}, \bibinfo{author}{Darlak, R.~S.} \&
  \bibinfo{author}{Tsai, L.~L.}
\newblock \bibinfo{journal}{\bibinfo{title}{Optical properties of
  hexahelicene}}.
\newblock {\emph{\JournalTitle{Journal of the American Chemical Society}}}
  \textbf{\bibinfo{volume}{89}}, \bibinfo{pages}{6191--6193}
  (\bibinfo{year}{1967}).

\bibitem{martin19681}
\bibinfo{author}{Martin, R.}, \bibinfo{author}{Flammang-Barbieux, M.},
  \bibinfo{author}{Cosyn, J.} \& \bibinfo{author}{Gelbcke, M.}
\newblock \bibinfo{journal}{\bibinfo{title}{1-synthesis of octa-and
  nonahelicenes. 2-new syntheses of hexa-and heptahelicenes. 3-optical rotation
  and ord of heptahelicene.}}
\newblock {\emph{\JournalTitle{Tetrahedron Letters}}}
  \textbf{\bibinfo{volume}{9}}, \bibinfo{pages}{3507--3510}
  (\bibinfo{year}{1968}).

\bibitem{FLAMMANGBARBIEUX1967743}
\bibinfo{author}{Flammang-Barbieux, M.}, \bibinfo{author}{Nasielski, J.} \&
  \bibinfo{author}{Martin, R.}
\newblock \bibinfo{journal}{\bibinfo{title}{Synthesis of heptahelicene (1)
  benzo [c] phenanthro [4, 3-g ]phenanthrene.}}
\newblock {\emph{\JournalTitle{Tetrahedron Letters}}}
  \textbf{\bibinfo{volume}{8}}, \bibinfo{pages}{743 -- 744}
  (\bibinfo{year}{1967}).

\bibitem{buss1996electronic}
\bibinfo{author}{Buss, V.} \& \bibinfo{author}{Kolster, K.}
\newblock \bibinfo{journal}{\bibinfo{title}{Electronic structure calculations
  on helicenes. concerning the chirality of helically twisted aromatic
  systems}}.
\newblock {\emph{\JournalTitle{Chemical physics}}}
  \textbf{\bibinfo{volume}{203}}, \bibinfo{pages}{309--316}
  (\bibinfo{year}{1996}).

\bibitem{vander1968fluorescence}
\bibinfo{author}{Vander~Donckt, E.}, \bibinfo{author}{Nasielski, J.},
  \bibinfo{author}{Greenleaf, J.} \& \bibinfo{author}{Birks, J.}
\newblock \bibinfo{journal}{\bibinfo{title}{Fluorescence of the helicenes}}.
\newblock {\emph{\JournalTitle{Chemical Physics Letters}}}
  \textbf{\bibinfo{volume}{2}}, \bibinfo{pages}{409--410}
  (\bibinfo{year}{1968}).

\bibitem{weigang1968low}
\bibinfo{author}{Weigang~Jr, O.} \& \bibinfo{author}{Dodson, P.~T.}
\newblock \bibinfo{journal}{\bibinfo{title}{Low-temperature circular dichroism
  of hexahelicene}}.
\newblock {\emph{\JournalTitle{The Journal of Chemical Physics}}}
  \textbf{\bibinfo{volume}{49}}, \bibinfo{pages}{4248--4250}
  (\bibinfo{year}{1968}).

\bibitem{furche2000circular}
\bibinfo{author}{Furche, F.} \emph{et~al.}
\newblock \bibinfo{journal}{\bibinfo{title}{Circular dichroism of helicenes
  investigated by time-dependent density functional theory}}.
\newblock {\emph{\JournalTitle{Journal of the American Chemical Society}}}
  \textbf{\bibinfo{volume}{122}}, \bibinfo{pages}{1717--1724}
  (\bibinfo{year}{2000}).

\bibitem{nakai2012theoretical}
\bibinfo{author}{Nakai, Y.}, \bibinfo{author}{Mori, T.} \&
  \bibinfo{author}{Inoue, Y.}
\newblock \bibinfo{journal}{\bibinfo{title}{Theoretical and experimental
  studies on circular dichroism of carbo [n] helicenes}}.
\newblock {\emph{\JournalTitle{The Journal of Physical Chemistry A}}}
  \textbf{\bibinfo{volume}{116}}, \bibinfo{pages}{7372--7385}
  (\bibinfo{year}{2012}).

\bibitem{farajian2003electronic}
\bibinfo{author}{Farajian, A.~A.}, \bibinfo{author}{Yakobson, B.~I.},
  \bibinfo{author}{Mizuseki, H.} \& \bibinfo{author}{Kawazoe, Y.}
\newblock \bibinfo{journal}{\bibinfo{title}{Electronic transport through bent
  carbon nanotubes: nanoelectromechanical sensors and switches}}.
\newblock {\emph{\JournalTitle{Physical Review B}}}
  \textbf{\bibinfo{volume}{67}}, \bibinfo{pages}{205423}
  (\bibinfo{year}{2003}).

\bibitem{accelrys2010materials}
\bibinfo{author}{Accelrys}.
\newblock \bibinfo{title}{Inc., materials studio software}
  (\bibinfo{year}{2010}).

\bibitem{vwn}
\bibinfo{author}{Vosko, S.~H.}, \bibinfo{author}{Wilk, L.} \&
  \bibinfo{author}{Nusair, M.}
\newblock \bibinfo{journal}{\bibinfo{title}{Accurate spin-dependent electron
  liquid correlation energies for local spin density calculations: a critical
  analysis}}.
\newblock {\emph{\JournalTitle{Canadian Journal of physics}}}
  \textbf{\bibinfo{volume}{58}}, \bibinfo{pages}{1200--1211}
  (\bibinfo{year}{1980}).

\bibitem{stratmann1998efficient}
\bibinfo{author}{Stratmann, R.~E.}, \bibinfo{author}{Scuseria, G.~E.} \&
  \bibinfo{author}{Frisch, M.~J.}
\newblock \bibinfo{journal}{\bibinfo{title}{An efficient implementation of
  time-dependent density-functional theory for the calculation of excitation
  energies of large molecules}}.
\newblock {\emph{\JournalTitle{The Journal of chemical physics}}}
  \textbf{\bibinfo{volume}{109}}, \bibinfo{pages}{8218--8224}
  (\bibinfo{year}{1998}).

\bibitem{cardia2014effects}
\bibinfo{author}{Cardia, R.}, \bibinfo{author}{Malloci, G.},
  \bibinfo{author}{Mattoni, A.} \& \bibinfo{author}{Cappellini, G.}
\newblock \bibinfo{journal}{\bibinfo{title}{Effects of tips-functionalization
  and perhalogenation on the electronic, optical, and transport properties of
  angular and compact dibenzochrysene}}.
\newblock {\emph{\JournalTitle{The Journal of Physical Chemistry A}}}
  \textbf{\bibinfo{volume}{118}}, \bibinfo{pages}{5170--5177}
  (\bibinfo{year}{2014}).

\bibitem{cardia2016electronic}
\bibinfo{author}{Cardia, R.} \emph{et~al.}
\newblock \bibinfo{journal}{\bibinfo{title}{Electronic and optical properties
  of hexathiapentacene in the gas and crystal phases}}.
\newblock {\emph{\JournalTitle{Physical Review B}}}
  \textbf{\bibinfo{volume}{93}}, \bibinfo{pages}{235132}
  (\bibinfo{year}{2016}).

\bibitem{dardenne2017tuning}
\bibinfo{author}{Dardenne, N.} \emph{et~al.}
\newblock \bibinfo{journal}{\bibinfo{title}{Tuning optical properties of
  dibenzochrysenes by functionalization: A many-body perturbation theory
  study}}.
\newblock {\emph{\JournalTitle{The Journal of Physical Chemistry C}}}
  \textbf{\bibinfo{volume}{121}}, \bibinfo{pages}{24480--24488}
  (\bibinfo{year}{2017}).

\bibitem{martin1996vibrational}
\bibinfo{author}{Martin, J.~M.}
\newblock \bibinfo{journal}{\bibinfo{title}{The vibrational spectra of
  corannulene and coronene. a density functional study}}.
\newblock {\emph{\JournalTitle{Chemical physics letters}}}
  \textbf{\bibinfo{volume}{262}}, \bibinfo{pages}{97--104}
  (\bibinfo{year}{1996}).

\bibitem{schaack2019helicene}
\bibinfo{author}{Schaack, C.} \emph{et~al.}
\newblock \bibinfo{journal}{\bibinfo{title}{Helicene monomers and dimers:
  chiral chromophores featuring strong circularly polarized luminescence}}.
\newblock {\emph{\JournalTitle{Chemistry--A European Journal}}}
  \textbf{\bibinfo{volume}{25}}, \bibinfo{pages}{8003--8007}
  (\bibinfo{year}{2019}).

\end{thebibliography}

\section*{Acknowledgements}

This research was supported in part by the Air Force Research Laboratory Materials and Manufacturing Directorate, through the Air Force Office of Scientific Research Summer Faculty Fellowship Program, Contract Numbers FA8750-15-3-6003 and FA9550-15-0001. This research was supported in part through computational resources and services provided by Advanced Research Computing at the University of Michigan, Ann Arbor. 

\section*{Author contributions statement}

V.V., D.S. and R.M. conceived the research idea(s),  V.S. conducted the simulations(s), V.V., V.S. and D.S. analysed the results.  All authors reviewed the manuscript. 

\section*{Competing interests}
The author(s) declare no competing interests.

\clearpage
\renewcommand\thefigure{S\arabic{figure}}   
\section*{Supplementary Data}

\setcounter{figure}{0}    

A decrease in the HOMO-LUMO gap implies better conjugation of bonds. However, one would expect bond lengths to increase during tensile straining greater than 25\% lowering the strength of bonds and affecting conjugation in the structure. In Figure S1, alternating inner helix bonds are indeed highly deformed during tension, with a few bonds stretching towards the $sp^3$ bond failure length of 1.54 \angstrom (at 100\% strain) making conjugation along the inner helix unlikely. 

\begin{figure}[h]
\centerline{\includegraphics[width=1.0\textwidth]{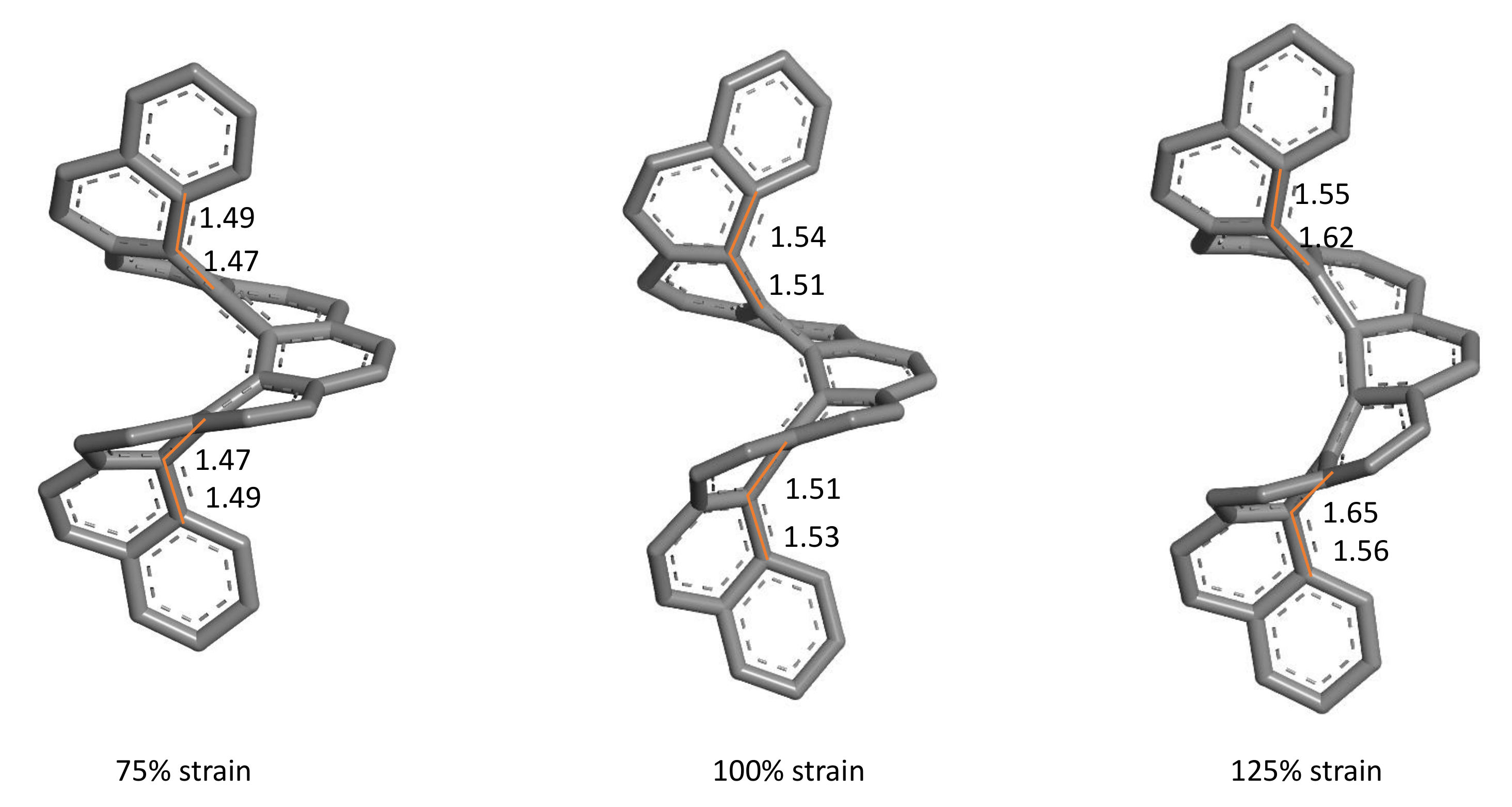}}
\centerline{(a)\hspace{3.5 in} (b)}
\caption{Failure strain in [9]-helicene: Bond lengths in the inner helix upon stretching. Close to 100\% strain, a bond length of 1.54~\angstrom (failure point of sp3 bonds) is reached.}
\label{s1}
\end{figure}

\end{document}